\documentstyle[12pt,psfig]{article}
\newcommand \beq{\begin{eqnarray}}
\newcommand \eeq{\end{eqnarray}}
\newcommand \ga{\raisebox{-.5ex}{$\stackrel{>}{\sim}$}}
\newcommand \la{\raisebox{-.5ex}{$\stackrel{<}{\sim}$}}
\newcommand{\doe}{This work was supported by the Director, Office of Energy 
                  Research, Division of Nuclear Physics of the Office of High 
                  Energy and Nuclear Physics of the U.S. Department of Energy 
                  under Contract No. DE-AC03-76SF00098.}
\begin{document}
\begin{flushright}
{\large LBL-37688}
\end{flushright}
\renewcommand{\thefootnote}{\fnsymbol{footnote}}
\setcounter{footnote}{0}
\begin{center}
{\large\bf Thermal Equilibration in an Expanding Parton Plasma\footnote{\doe}
}\\
\vspace{1.5cm}
{\bf H. Heiselberg}\\
{\it NORDITA, Blegdamsvej 17, DK-2100 Copenhagen \O., Denmark}\\
\vspace{0.5cm}
{\bf Xin-Nian Wang}\\
{\it Nuclear Science Division, Mailstop 70A-3307}\\
{\it Lawrence Berkeley National Laboratory}\\
{\it University of California, Berkeley, CA 94720 USA}\\
\end{center}
\date{May, 1995}

\vskip 6\baselineskip

\begin{abstract}
\baselineskip=15pt

Thermalization in an expanding parton plasma is studied within
the framework of Boltzmann equation in the absence of any mean fields.
In particular, we study the time-dependence of the relaxation time
to the lowest order in finite temperature QCD and how such time-dependence
affects the thermalization of an expanding parton plasma. Because of 
Debye screening and Landau damping at finite temperature, the relaxation 
time (or transport rates) is free of infrared divergencies in both 
longitudinal and transverse interactions. The resultant relaxation time
decreases with time in an expanding plasma like $1/\tau^\beta$,
with $\beta<1$. We prove in this case that thermal equilibrium
will eventually be established given a long life-time of the system. 
However, a fixed momentum cut-off in the calculation of the relaxation 
time gives rise to a much stronger time dependence which will slow 
down thermal equilibrium. It is also demonstrated that the ``memory effect'' 
of the initial condition affects the approach to thermal equilibrium and 
the final entropy production.
\end{abstract}

\vfill
\noindent PACS numbers:12.38.Mh,25.75+r,12.38.Bx

\newpage

\baselineskip=16pt
\section{Introduction} 

Perturbative-QCD-based models developed in the last few years 
predict that nucleus-nucleus collisions at future collider
energies are dominated by hard or semihard processes 
\cite{JBAM,KLL,HIJING,PCM}. These processes happen during 
the very early stage of the collisions and they produce a 
rather large number of semihard partons which essentially 
form a hot and undersaturated parton gas \cite{PCM,KEXW93}. 
However, this parton gas is initially far away from thermal 
and chemical equilibrium \cite{BMTW}. Secondary
parton scatterings in the gas may eventually lead to
local thermal and chemical equilibrium if the parton
interactions are sufficiently strong.

Transport calculations based on a semiclassical parton 
cascade model~\cite{PCM} indicate that
thermal equilibrium could be established within  a rather
short time of about 1 fm/$c$.  However, the complexity of
the Monte Carlo simulations makes it difficult to obtain
a lucid understanding of the dependence of the thermalization 
time on the many parameters employed in the model.  One such
parameter is the cut-off of momentum transfer in binary
parton scatterings. The cut-off was first introduced to regularize 
the infrared divergency of the cross section between two massless
partons in high-energy $pp$ and $p\bar{p}$ collisions \cite{XW91}.
The value of the momentum cut-off is determined phenomenologically
to reproduce the measured total cross sections of $pp$ and $p\bar{p}$
collisions. However, this cut-off is not necessary anymore in a 
high-temperature quark-gluon plasma, since the Debye screening and 
Landau damping provide natural regularizations of the infrared 
divergency. Since transport times depend sensitively on the screening
masses which in turn depend on the temperature, the introduction
of an artificial cut-off could give rise to a completely different
behavior of the thermalization time and consequently the approach to 
thermal equilibrium.

The approach to thermal equilibrium in relativistic heavy ion
collisions is dictated by the competition between expansion and
parton interactions \cite{HHXW95}. If the expansion is much rapid than the
typical collision time among partons, e.g., shortly after partons
are initially produced, the expansion is closer to free-streaming
than hydrodynamic expansion. Only at times in the order of the
collision time may the parton gas reach local thermal equilibrium
and expand hydrodynamically. Furthermore, the time dependence of 
the collision time (or the relaxation time) will determine whether
the system can eventually reach local thermal equilibrium because 
of the competition between expansion and parton interactions. If
the collision time increases rapidly with time, the parton system
may never thermalize, leading only to a free-streaming limit.
The collision time, therefore, is a very important quantity which
in turn depends sensitively on the infrared behavior of parton
interactions.

In QCD, parton scattering cross sections exhibit a quadratic 
infrared singularity
due to the exchange of a massless gluon. The infrared behavior can
be improved by including corrections from hard thermal loops to the
gluon propagators. Resummation of these thermal loops gives rise
to an effective gluon propagator which screens long range 
interactions (Debye screening). Braaten and Pisarski~\cite{BRPIRS}
have developed this resummation technique systematically and
used it to calculate the damping rate of a soft gluon ($p\sim gT$)
which is gauge invariant and complete to the leading order in the
QCD coupling constant $g$ \cite{BRPI90}. For a fast particle ($p\ga T$), 
the  exchanged gluons probe the static limit of the magnetic interactions
which by the transversality condition are not screened. One thus
has to introduce a nonperturbative magnetic screening mass to
regularize the logarithmic infrared singularity in the static 
limit \cite{CBAM92,HHCP93,PISA93}. As a result, the damping rate 
for an energetic particle to the leading order in $g$ is
\begin{equation}
  \Gamma \sim T \left[ \alpha_s \ln(1/\alpha_s)+{\cal O}(\alpha_s)\right]\; , 
  \label{eq:damp1}
\end{equation}
where $\alpha_s=g^2/4\pi$. However, as we will argue, damping 
rates do not determine how fast a system approach local thermal 
equilibrium. What really determine the thermalization processes 
are the transport rates which are free of the logarithmic
divergency after the resummation of thermal loops
\cite{DG85,HK85}. This is because thermalization is achieved 
to the leading order mainly through momentum changes in elastic 
scatterings. Thus, the effective cross section should be weighted 
by the momentum transfer and the dynamic screening due to the 
Landau damping of the gluons is sufficient to regularize the 
logarithmic singularity in the transverse interactions 
\cite{BP90,HH94,THOM94}. The resultant transport times for a 
system near thermal equilibrium behave like
\begin{equation}
\frac{1}{\tau_{\rm tr}}\sim T \alpha_s^2\ln(1/\alpha_s) \label{eq:trans1}
\end{equation}
to the leading order in $\alpha_s$.

{}For a system near local thermal equilibrium, the time dependence 
of the transport times is through the temperature according to 
Eq.~(\ref{eq:trans1}). This dependence is in general slower 
than $1/\tau$ and thus can lead to local thermal equilibrium 
according to our earlier argument based on the relaxation time 
approximation \cite{HHXW95}. However, if one introduces an 
artificial cut-off for the momentum transfers of elastic parton 
scatterings as in the numerical simulation of a classical 
parton cascade~\cite{PCM}, the time dependence will be much stronger.
Consequently, as we will demonstrate in this paper, the system will 
approach local thermal equilibrium much slowly. We will also demonstrate
that inclusion of the screening effects is the key to a
slower time dependence of the relaxation time, therefore a faster
approach to thermal equilibrium.

This paper is organized as follows. In Section II, we first re-examine 
the Boltzmann equation and the evaluation of the damping rate and the 
relaxation time to the lowest order, including only $2\leftrightarrow 2$ 
processes. We will also discuss
the time dependence of the relaxation time in different
scenarios. In Section III, we will solve the Boltzmann
equation in the relaxation time approximation and demonstrate
how time dependence of the relaxation time will
affect the approach to thermal equilibrium. We also
show how initial conditions of a system affect
the thermalization processes and the final total
entropy production (or ``memory effect'') in Section IV.
Finally in Section V we give a summary and an outlook,
especially of the numerical simulations of parton
thermalization, taking into account of the Debye
screening and Landau damping effects without double counting.

\section{Time dependence of thermalization time}

In a system with two-components as, e.g., the  quark and gluon plasma,
the one that interacts the strongest will thermalize faster than the
other.  Subsequently there will be momentum and energy transfer 
between them. Since numerical simulations indicate that the initially
produced partons are mostly gluon, we consider here a gluon gas only
for simplicity. The Debye screening of color fields in the
presence of semihard gluons \cite{BMW} will also allow us to
neglect the effect of mean fields.  We furthermore assume that the spatial 
variation of the system is small on the scale of a collision length 
so that we can approximate the evolution of the system by 
the Boltzmann equation \cite{GROOT},
\begin{eqnarray}
  v_1\!\cdot\!\partial f_1(p_1)&=&-\nu_2\int dp_2dp_3dp_4
  F_{1234}[f]\frac{1}{2}|M_{12\to 34}|^2 (2\pi)^4\delta^4(P_1+P_2-P_3-P_4), 
  \label{eq:boltz1} \\
  F_{1234}[f]&=&f_1f_2(1\pm f_3)(1\pm f_4)-f_3f_4(1\pm f_1)(1\pm f_2),
\end{eqnarray}
where $P_i=(|p_i|,{\bf p}_i)$ are the four-momenta of massless partons
and $dp_i\equiv d^3p/(2\pi)^3$. To keep the formula general, $\pm$
are used for bosons (gluons) and fermions (quarks and anti-quarks),
respectively. The statistical factor $\nu_2$ is $2(N_c^2-1)=16$ for 
gluons and $12N_f$ for $N_f$ flavors of quarks and anti-quarks with
$N_c=3$ colors. The squared
matrix element,  $|M_{12\to 34}|^2\equiv |{\cal M}_{12\to 34}|^2/
(16E_1 E_2 E_3 E_4)$ is summed over final states and averaged over
initial states. For gluon-gluon scatterings,
\begin{equation}
|{\cal M}_{12\to 34}|^2 = C_{gg}4g^4\left(3-\frac{su}{t^2}-\frac{st}{u^2}
  -\frac{tu}{s^2}\right), \label{eq:cross}
\end{equation}
where $C_{gg}=N_c^2/(N_c^2-1)=9/8$ is the color factor of gluon-gluon
scatterings, $s$, $t$ and $u$ are the Mandelstam variables. There is
clearly a quadratic singularity for small energy $\omega$ and
momentum $q$ transfers because of the long range interactions mediated
by the massless gauge bosons. Because the final state has two identical
particles, $su/t^2$ should contribute equally as $st/u^2$ in 
Eq.~(\ref{eq:cross}). We thus can approximate
$|{\cal M}_{12\to 34}|^2 \approx 8 C_{gg} g^4 s^2/t^2$ for
small-angle gluon scatterings.  Since the collisional
integral is dominated by contributions from near the singularity,
we can assume a small angle scattering approximation, i.e.,
$\omega, q \ll E_1, E_2$. Then energy-momentum conservation leads to
\begin{eqnarray}
  {\bf p}_3&=&{\bf p}_1+{\bf q},\quad {\bf p}_4={\bf p}_2-{\bf q}
\;, \nonumber \\
  E_3&=& E_1+\omega,\quad  E_4= E_2-\omega \;, \nonumber \\
  \omega&\approx& {\bf v_1}\!\cdot\!{\bf q}\approx{\bf v_2}\!\cdot\!{\bf q}\; .
  \label{eq:consv}
\end{eqnarray}
The integration over $p_3$ and $p_4$ can be rewritten as
\begin{equation}
(2\pi)^4\int dp_3\,dp_4\delta^4(P_1+P_2-P_3-P_4)
=\frac{1}{(2\pi)^2}\int d^3q\int_{-q}^{q} d\omega 
\delta(\omega-{\bf v_1}\!\cdot\!{\bf q})
\delta(\omega-{\bf v_2}\!\cdot\!{\bf q})\; .
\end{equation}

In a medium, one can use a resummation technique to include 
an infinite number of loop corrections to
the gluon exchange. Using Dyson's equation, this amounts to
an effective gluon propagator. One can use this effective propagator
to obtain the effective matrix element squared for forward
gluon scatterings (see Appendix A),
\begin{equation}
  |M_{gg}|^2 \approx 2 C_{gg} g^4 \left|\frac{1}{q^2+\pi_L(x)}
  -\frac{(1-x^2)\cos\phi}
  {q^2(1-x^2)+\pi_T(x)}\right|^2\; ,
  \label{eq:matrix1}
\end{equation}
where $\cos\phi=
({\bf v}_1\times\hat{\bf q})\!\cdot\!({\bf v}_1\times\hat{\bf q})$
and $x=\omega/q$. The scaled self-energies in the
long wavelength limit are given by \cite{WELD82}
\begin{eqnarray}
  \pi_L(x)&=&q^2_D\left[1-\frac{x}{2}
  \ln\left(\frac{1+x}{1-x}\right) + i \frac{\pi}{2}x \right] 
  \, , \label{eq:self1}\\
  \pi_T(x)&=&q^2_D\left[\frac{x^2}{2}+\frac{x}{4}(1-x^2)
  \ln\left(\frac{1+x}{1-x}\right) - i \frac{\pi}{4}x(1-x^2)\right]
  \, ,\label{eq:self2}
\end{eqnarray}
where $q_D^2=g^2(N_c+N_f/2)T^2/3$ is the Debye screening mass in 
thermal QCD. The imaginary parts provide Landau damping to
parton interactions in a thermal medium. We see that the longitudinal 
interactions are screened 
by thermal interactions. However, the transverse interactions still 
have a logarithmic singularity in the static limit. This singularity
can only be regularized by introducing a nonperturbative magnetic
screening mass in the calculation of the damping rate of a
fast parton.

We can use the definition of parton interaction rates,
\begin{eqnarray}
\Gamma_{\rm loss}(p_1)&=&\frac{\nu_2}{(2\pi)^5}\int d^3p_2 d^3q d\omega
f_2(1\pm f_3)(1\pm f_4)\frac{1}{2}|M_{12\to 34}|^2
\delta(\omega-{\bf v_1}\!\cdot\!{\bf q})
\delta(\omega-{\bf v_2}\!\cdot\!{\bf q})\; ,\label{eq:rate1}\\
\Gamma_{\rm gain}(p_1)&=&\frac{\nu_2}{(2\pi)^5}\int d^3p_2\,d^3qd\omega
f_3f_4(1\pm f_2)\frac{1}{2}|M_{12\to 34}|^2
\delta(\omega-{\bf v_1}\!\cdot\!{\bf q})
\delta(\omega-{\bf v_2}\!\cdot\!{\bf q})\; , \label{eq:rate2}
\end{eqnarray}
and rewrite the Boltzmann equation as
\begin{eqnarray}
  v_1\!\cdot\!\partial f_1&=&-f_1\Gamma_{\rm loss}(p_1)
  +(1\pm f_1)\Gamma_{\rm gain}(p_1) \nonumber \\
  &=&-\Gamma(p_1) (f_1-\tilde{f})\; , \label{eq:boltz2}
\end{eqnarray}
where
\begin{equation}
  \Gamma(p)=\Gamma_{\rm loss}(p)\mp\Gamma_{\rm gain}(p)
\end{equation}
is usually referred to as the damping rate of a particle 
(or a quasiparticle), and $\tilde{f}$ is defined as
\begin{equation}
  \tilde{f}(p)=\frac{\Gamma_{\rm gain}(p)}
  {\Gamma_{\rm loss}(p)\mp \Gamma_{\rm gain}(p)}\; .
\end{equation}
For a system in local thermal equilibrium, one can relate 
the damping rate to the imaginary part of the gluon 
self-energy \cite{WELD83},
\begin{equation}
{\rm Im}\Pi^{\mu}_{\;\mu}(p)=-2 (p\!\cdot\! u)\Gamma(p)\; ,
\end{equation}
where the factor 2 comes from our definition of the interaction
rates among identical particles.
One can also show that, if $f(p)$ takes the local equilibrium
form $f^{\rm eq}(p)=(\exp(p\cdot u/T)\mp 1)^{-1}$,
\begin{equation}
  \frac{\Gamma_{\rm loss}(p)}{\Gamma_{\rm gain}(p)}=e^{p\cdot u/T},
\end{equation}
using the energy and momentum conservation and the identity
$1\pm f^{\rm eq}(p)=f^{\rm eq}(p)\exp(p\!\cdot\! u/T)$. Therefore,
by definition, $\tilde{f}(p)$ becomes $f^{\rm eq}(p)$.
Thus, the global equilibrium distribution $f^{\rm eq}(p)$ is 
a solution to the Boltzmann equation if the flow velocity $u$ is 
independent of space and time.

We can complete the angular integrations in Eqs.~(\ref{eq:rate1})
and (\ref{eq:rate2}). Making approximations $f(p_3)\approx f(p_1)$
and $f(p_4)\approx f(p_2)$, we can also complete the integration 
over $p_2$ by cutting off the integration over $q$ at 
$q_{\rm max}\approx 3T$ \cite{HH94}. We then obtain the gluon
damping rate,
\begin{equation}
  \Gamma^{gg}=\frac{g^4}{4\pi}\frac{C_{gg}\nu_g}{12} T^3
  \int_{-1}^1dx\int_0^{q^2_{\rm max}}dq^2
\left\{\frac{1}{|q^2+\pi_L(x)|^2}+
\frac{1}{2}\frac{(1-x^2)^2}{|q^2(1-x^2)+\pi_T(x)|^2}\right\}\; ,
\end{equation}
including only gluon-gluon scatterings.

The contribution from longitudinal interactions is finite
and proportional to $g^2 T$ due to the Debye screening. However,
Debye screening is absent in the transverse interactions in the
static limit. There is a logarithmic divergency even if Landau 
damping is taken into account. One solution to this problem is
to add a nonperturbative magnetic mass $m_{mag}\sim g^2T$
to the transverse self-energy $\pi_T(x)$. In this case, the
dominant contribution of the transverse interactions comes
from $m_{\rm mag}\la q \la q_D$ and thus  is 
proportional to $g^2T\ln(q_D/m_{\rm mag})$
which is independent of $q_{\rm max}\gg q_D$. With $\pi_L(x)$
and $\pi_T(x)$ given by Eqs.~(\ref{eq:self1}) and (\ref{eq:self2}),
we can complete the numerical integration. A fit to the 
numerical result gives us
\begin{equation}
  \Gamma^{gg}=\frac{g^4}{4\pi}\frac{C_{gg}\nu_g}{6}\frac{T^3}{q_D^2}
  \left[(\ln\frac{q_D^2}{m_{\rm mag}^2}-1.0
  +2.0\frac{m_{\rm mag}^2}{q_D^2} - 0.32\frac{q_D^2}{q_{\rm max}^2})
  + 1.1\frac{q_{\rm max}^2}{q_{\rm max}^2+q_D^2}\right]\; ,
  \label{eq:19}
\end{equation}
where the first term comes from the transverse interactions while
the second from the longitudinal ones.
Using the estimate of $m_{\rm mag}\approx 0.255\sqrt{N_c/2}g^2T$
from Ref.~\cite{TBBM93} and neglecting the quark contribution to
the Debye screening mass, we have
\begin{equation}
  \Gamma^{gg}\approx N_c\alpha_s T
  \left[\ln(1/\alpha_s) -0.1+{\cal O}(\alpha_s) \right] \;. \label{eq:20}
\end{equation}
Note that contributions to the order $\alpha_s^2$ in Eq.~(\ref{eq:19})
have been neglected, since they are not complete in our calculation.
In order to have a complete calculation of such higher order 
corrections, one has to include thermal vertex and vacuum corrections 
which should depend on the renormalization scale. The final result 
to this order should be invariant under the renormalization group.
This result agrees with previous calculations \cite{CBAM92,HHCP93,PISA93}
to the leading order of $\alpha_s$ which depends only on the
imaginary part of the transverse self-energy,
$\pi_{T}(x)\approx -i(\pi/4)q_D^2 x$ at small $x$. Inclusion of 
the full expression of the self-energy only contribute to the
next order corrections. The increase in scattering by including quarks  
is exactly compensated by the increase in Debye screening due to 
quarks \cite{HHCP93}.

For a soft gluon ($p\sim gT$), one can not neglect its thermal
mass anymore. The damping rate for a gluon at rest will
not have the logarithmic divergency in the transverse interaction,
since the exchanged gluon must carry nonzero momentum and energy
at least of order of $gT$ and thus never approach to the static limit.
In addition the transverse (magnetic) interactions are reduced by 
velocity factors which for the massive partons are smaller than the 
speed of light.
The damping rate in this case was found by Braaten and Pisarski
\cite{BRPI90} to the leading order as
\begin{equation}
  \Gamma^{gg}(0)\approx 1.1 N_c \alpha_s T\; .
\end{equation}
Apparently, the damping rate has a nontrivial momentum 
dependence \cite{PISA89}.

As we have mentioned, $f^{\rm eq}(p\!\cdot\!u/T)$ is a solution to
the Boltzmann equation as far as the flow velocity is uniform in
space and time. We should emphasize here that the damping rate 
{\it does not} determine how rapidly a system near equilibrium
approaches it as one would naively think. The thermalization time 
is actually related to the transport rates 
\cite{DG85,HK85,BP90,HH94,THOM94}. The easiest
way to prove this is to check that the logarithmic divergency
that has plagued the calculation of the damping rate of a fast gluon
does not appear in the Boltzmann equation. To check this, we make
the following expansion:
\begin{equation}
  f(p_{3,4})\approx f(p_{1,2})\pm \omega f'(p_{1,2})
  +\frac{\omega^2}{2}f''(p_{1,2})\; ,
\end{equation}
for small angle scatterings. The function $F_{1234}[f]$ in 
Boltzmann equation becomes
\begin{equation}
  F_{1234}[f]=-\frac{q^2x^2}{2}\left[f_1(1\pm f_1)f''_2
  +f_2(1\pm f_2)f''_1 -2(1\pm f_1 \pm f_2)f'_1f'_2\right] \; ,
\end{equation}
which is proportional to $q^2$. Here we have dropped terms 
linear in $x$ since they vanish after integration over $x$.
One can verify that this function after expansion still
vanishes for the equilibrium distribution, $F_{1234}[f^{\rm eq}]=0$.
For a system away from equilibrium, the collisional integral
in the Boltzmann equation is nonzero but finite despite
the logarithmic singularity in the transverse part of the
matrix element squared, $|M_{12\to 34}|^2$ when $q, \omega \to 0$. 
Because of the factor $q^2$ in $F_{1234}[f]$, Landau damping 
in the self-energy of the exchanged gluon is sufficient 
to give a finite value of the collisional integral. In other words,
thermalization not only depends on the parton interaction rates
but also on the efficiency of transferring momentum in each
interaction. Those interactions with zero energy and momentum
transfers do not contribute to the thermalization process, though
their cross sections are infinitely large. This is why the
thermalization time and other transport coefficients do 
not suffer from the infrared
divergency as pointed out in a number of papers 
\cite{HHCP93,BP90,HH94,THOM94}.

For a system near equilibrium, one can characterize the
deviation from equilibrium by $\delta f=f-f^{\rm eq}
=-\theta(p)p\!\cdot\!\partial f/(p\!\cdot\!u)$
in a relaxation time approximation. Equivalently, one has
\begin{equation}
  \frac{1}{p\!\cdot\!u}p\!\cdot\!\partial f
=-\frac{f-f^{\rm eq}}{\theta(p)} \; . \label{eq:boltz3}
\end{equation}
In general the relaxation time $\theta(p)$ depends
on momentum $p$ and in principle can be obtained by solving
the linearized Boltzmann equation. In this paper, we neglect the
momentum dependence of the relaxation time. For a pure gluonic
gas near local thermal equilibrium where thermalization is 
achieved through viscous relaxation, the relaxation time 
is (Appendix B)
\begin{equation}
  \frac{1}{\theta}\simeq 0.92 N_c^2 T\alpha_s^2
  \ln\left(\frac{1.6}{N_c\alpha_s}\right)\; .
\end{equation}
Again, because of the extra factor $q^2$ appearing in the transport
rate, the Debye screening and Landau damping are sufficient to
regularize the effective transport cross section. The dominant
contribution comes from interactions with $q_D \la q\la q_{\rm max}$,
leading to a logarithmic factor $\ln(q_{\rm max}/q_D)$ as compared
to $\ln(q_D/m_{\rm mag})$ in the gluon damping rate. Therefore,
the dependence of the relaxation time on the (weak) coupling 
constant and the color dimension $N_c$ is quite different from the 
gluon damping rate.

If the system is close to thermal equilibrium, the hydrodynamic
equations from energy-momentum conservation to the zeroth order 
of $\delta f$ can give us the time evolution of the temperature
$T$. For an ideal gluon gas with one-dimensional expansion, $T$
decreases like $T/T_0=(\tau_0/\tau)^{1/3}$. Therefore, the relaxation
time $\theta$ increases with time with a power of $1/3$,
\begin{equation}
  \theta=\frac{(\tau/\tau_0)^{1/3}}{0.92 N_c^2 T_0 \alpha_s^2
    \ln(1.6/N_c\alpha_s)} \;. \label{eq:relax1}
\end{equation}
A more general time dependence of the relaxation time can
have a power-law form,
\begin{equation}
  \theta=\theta_0 (\tau/\tau_0)^{\beta}\; , \label{eq:relax2}
\end{equation}
which also covers both the constant ($\beta=0$) and the linear 
($\beta=1$) cases as have been studied by several authors 
\cite{BAYM,KM85,BBR,GAV91}. The latter case arises when a 
constant scattering cross section, $\sigma$,
is assumed for the relaxation time, $\theta\sim (\sigma n)^{-1}$, 
and with a density decreasing as $n\sim \tau^{-1}$ due to 
one-dimensional expansion. For a system far away from thermal 
equilibrium, the time dependence may differ from Eq.~(\ref{eq:relax1}).
Our earlier calculations \cite{HHXW95} show that an initially 
free-streaming system has only a logarithmic time dependence. 
This is because the phase space for small angle scatterings 
($q\sim q_D$) opens 
up quadratically with time in the free-streaming case and it 
balances the decrease in parton density.

We would like to emphasize that the weak time dependence of the
relaxation time in Eq.~(\ref{eq:relax1}) 
depends very sensitively on the Debye screening of the 
small-angle parton scatterings which restricts the momentum
transfer to $q\ga q_D=gT$. Smaller Debye screening mass due to
the decrease of the temperature, gives a larger interaction
rate which then compensates the decrease of the parton density
and thus gives the weak time dependence of the relaxation time.
If we use a fixed momentum cut-off $q_{\rm cut}$, as in most of 
the numerical simulations of parton production \cite{HIJING} 
and cascade \cite{PCM}, instead of a time dependent Debye 
screening mass, the effective transport cross section will 
remain constant, proportional
to $N_c^2\alpha_s^2/T^2\ln(T^2/q_{\rm cut}^2)$.
The resultant relaxation time 
\begin{equation}
 \frac{1}{\theta}\propto N_c^2\alpha_s^2 T\ln(T/q_{\rm cut})
\end{equation}
will increase more rapidly with time.

If we have to include transverse expansion later in the
evolution of a system, then the temperature will decrease
faster, like $\tau_0/\tau$, than in the one-dimensional case assuming 
hydrodynamic expansion. The relaxation time even with the 
inclusion of Debye screening will increase linearly with time. 
The relaxation time with a momentum cut-off applied to parton 
interactions will increase faster than linear with time, which will 
only lead the system into free-streaming.

 At this point we should emphasize that we have only 
considered the lowest order contribution from $2\leftrightarrow 2$ 
processes in our calculation of the relaxation time. In principle,
higher order processes, like $2\leftrightarrow 2+n$, should also 
contribute to the thermalization. Such processes can be included by
considering high order thermal vertex corrections. For a 
complete calculation, one should also include vacuum
corrections and the result should depend on the renormalization
scale and obey the renormalization group equation.
In general, contributions from $2\leftrightarrow 2+n$ processes 
should have a form \cite{field},
\begin{equation}
\Gamma_n \sim \Gamma_0[\alpha_s\ln(q^2/q_0^2)]^n,
\end{equation}
where $q^2$ is the momentum scale of these processes. 
At zero temperature, $q_0$ is some confinement scale below 
which perturbative QCD is no longer applicable. At finite 
temperature, $q_0$ is very likely to be replaced by screening 
masses. Since the largest momentum scale in a system at finite 
temperature is $q^2 \sim T^2$, the leading correction from
$2\leftrightarrow 2+n$ processes must be,
\begin{equation}
\Gamma_n \sim \Gamma_0 [\alpha_s\ln(1/\alpha_s)]^n.
\end{equation}
Such corrections therefore are high orders in
$\alpha_s\ln(1/\alpha_s)$ and are negligible in
the week couple limit.  For temperatures not far
above the QCD phase transition temperature $T_c\sim 200$ MeV,
the strong coupling constant is  not very small.
The above contributions might not be negligible. However,
for an order-of-magnitude estimate, we can neglect
these higher order contributions. If one considers
the chemical equilibration of a kinetically thermalized
system as in Ref.~\cite{BMTW},  $n \leftrightarrow m$ multiplication 
processes become very important. The leading contribution to
the chemical equilibration, in this case, comes from
$2\leftrightarrow 3$ processes.

\section{Approach to thermal equilibrium}
Let us consider the early stage of a very heavy ion collision
where transverse expansion is not important yet. We then can
treat the system as a one-dimensional system. We assume along 
with Bjorken \cite{BJ93} a scaling flow velocity
\begin{equation}
  u_{\mu}=\frac{x_{\mu}}{\tau}=(\cosh\eta,\sinh\eta,{\bf 0}_{\perp})
\end{equation}
in the longitudinal direction, where
\begin{equation}
  \tau=\sqrt{t^2-z^2}\;,\;\;\; 
  \eta=\frac{1}{2}\ln\left(\frac{t+z}{t-z}\right)
\end{equation}
are the proper time and spatial rapidity, respectively.
In terms of these new variables, the Boltzmann equation in 
the relaxation time approximation for a system near thermal
equilibrium becomes
\begin{equation}
  \frac{\partial f}{\partial \tau}-\frac{\tanh\xi}{\tau}
  \frac{\partial f}{\partial \xi}=-\frac{f-f^{\rm eq}}{\theta}\; ,
   \label{eq:boltz4}
\end{equation}
where $\xi=\eta-y$ and $y$ is the rapidity of a particle,
\begin{equation}
  y=\frac{1}{2}\ln\left(\frac{E+p_z}{E-p_z}\right)\; .
\end{equation}
Since $p\!\cdot\!u=p_T\cosh\xi$, we can see that the solution
to the above Boltzmann equation is a function of $\xi$, $\tau$ and
$p_T$, and is Lorentz invariant  under longitudinal boost.
One may then, as done in \cite{BAYM}, solve the Boltzmann equation in
the central slice only, i.e., $\eta=0$.
To simplify the problem, we assume the system is already in
chemical equilibrium so that the gluon chemical potential vanishes.
This is not always true in a realistic situation as shown by
numerical simulations \cite{KEXW93} of initial parton production at around
RHIC energies. However, at LHC energies, the small-$x$ behavior
of the parton distributions as measured by recent HERA 
experiments \cite{HERA} gives much higher densities of initially
produced partons very close to chemical
equilibrium \cite{EKR94}. For a given momentum and time
dependence of the relaxation time, one can find the solution
to Eq.~(\ref{eq:boltz4}) in an integral form,
\begin{eqnarray}
  f(p_T,\xi,\tau)&=&e^{\chi_0-\chi}f_0(p_T,\sinh^{-1}
  (\frac{\tau\sinh\xi}{\tau_0})) \nonumber \\
  &+&\int_{\chi_0}^{\chi}d\chi' e^{\chi-\chi'}
  f^{\rm eq}(p_T,\sinh^{-1}(\frac{\tau\sinh\xi}{\tau'}),T')\; .
  \label{eq:solu1}
\end{eqnarray}
Here $f_0(p_T,\xi)$ is the initial distribution at time $\tau_0$,
and the time dependence of the temperature in $f^{\rm eq}(p_T,\xi,T)$ 
is determined by requiring the energy density for $f$ and $f^{\rm eq}$
be equal at any time,
\begin{equation}
  \epsilon(T)\equiv\int dp E f(p)=\int dp E f^{\rm eq}(p,T) \; .
  \label{eq:solu3}
\end{equation}
The variable $\chi$ is defined by
\begin{equation}
  \chi(p_T,\xi,\tau)=\chi_0+\int_{\tau_0}^{\tau}d\tau'
  \theta^{-1}(p_T,\sinh^{-1}(\frac{\tau\sinh\xi}{\tau'}),\tau')\; .
  \label{eq:solu2}
\end{equation}
If we take the form in Eq.~(\ref{eq:relax2}) for a
momentum-independent relaxation time, then
\begin{equation}
  \chi=\chi_0\left(\frac{\tau}{\tau_0}\right)^{1-\beta},  \;\;
  \chi_0=\frac{\tau_0/\theta_0}{1-\beta} \, ,
\label{eq:chi}
\end{equation}
for $\beta\neq 1$. The case for $\beta=1$ was studied by Gavin 
\cite{GAV91}. For a more general discussion, let us also assume a
initial parton distribution
\begin{equation}
  \frac{dN}{d\xi dp_T^2}=w(p_T)\theta(|\xi|-Y)
\end{equation}
with a simple plateau distribution in $\xi$. One can also
consider a gaussian form of distribution \cite{PLBMXW}, but the
above form is simpler. The width $Y$ is
a measure of the initial correlation between space and momentum,
or $\eta$ and $y$. This width which normally depends on the 
transverse momentum $p_T$ \cite{ZLMG,PLBMXW} can be estimated by 
the uncertainty principle. We will assume it a constant for simplicity
and will study how the thermalization process depends on the
initial condition by varying $Y$.  Here, $w(p_T)$ is the $p_T$
distribution which usually has a power-law or exponential form.
The corresponding phase space density within a volume $V$ is
\begin{equation}
  f_0(p_T,\xi)=\frac{2(2\pi)^2}{\nu_gV}
  \frac{w(p_T)}{p_T\cosh\xi}\theta(|\xi|-Y) \; ,
  \label{eq:init1}
\end{equation}
with an initial energy density
\begin{equation}
  \epsilon_0=\frac{2\sinh Y}{V}\int dp_T^2 p_Tw(p_T)\; .
  \label{eq:init2}
\end{equation}

Following Baym \cite{BAYM}, we study the solution to 
the Boltzmann equation by taking the first moment 
(energy density) of Eq.~(\ref{eq:solu1}) with respect to 
the single parton energy. Defining,
\begin{equation}
  G=\frac{\tau\epsilon(\tau)}{\tau_0\epsilon_0}\; , \label{eq:G}
\end{equation}
we have from Eq.~(\ref{eq:solu1})
\begin{equation}
  G(\chi)=e^{\chi_0-\chi}H_Y((\chi_0/\chi)^{\frac{1}{1-\beta}})
  +\int_{\chi_0}^{\chi}d\chi' e^{\chi'-\chi}G(\chi')
  h((\chi'/\chi)^{\frac{1}{1-\beta}}) \; , \label{eq:solu4}
\end{equation}
where $\chi'/\chi=(\tau'/\tau)^{1-\beta}$ from Eq.~(\ref{eq:chi}), and
\begin{eqnarray}
  H_Y(a)&=&\frac{1}{\sinh Y}\int_0^{Y}d\xi \sqrt{1+a^2\sinh^2\xi}\; ,\\
  h(a)&=&\frac{1}{2}\left(a + 
  \frac{\arcsin\sqrt{1-a^2}}{\sqrt{1-a^2}}\right)\: .
\end{eqnarray}
If the initial distribution $f_0$ is an equilibrium one,
one should replace $H_Y(a)$ by $h(a)$ in Eq.~(\ref{eq:solu4}).
However, $H_Y(a)$ [with $H_Y(1)=1$ and 
$H_Y(0)=Y/\sinh Y$] covers 
a wide selection of initial distributions with different values of $Y$. 
{}For $Y=1.23$ (i.e., $Y/\sinh Y=h(0)=\pi/4$), $H_Y(a)$ is very 
similar to $h(a)$ for an isotropic initial distribution. 
When $Y=0$, the initial distribution corresponds to the Bjorken
scaling ansatz ($\eta=y$). One then has $H_0(a)=1$, which is similar
to the case discussed by Baym \cite{BAYM}. After performing a partial 
integration in Eq.~(\ref{eq:solu4}), we have finally
\begin{equation}
  \int_{\chi_0}^{\chi}d\chi' e^{\chi'}\frac{d}{d\chi'}
  [G(\chi')h((\chi'/\chi)^{\frac{1}{1-\beta}})]
  =e^{\chi_0}[H_Y(\tau_0/\tau)-h(\tau_0/\tau)] \; .
  \label{eq:solu5}
\end{equation}
The discussion of the condition for equilibrium at large times, $\tau$,
is now similar to our earlier analysis \cite{HHXW95} even 
with the general initial condition assumed. However, as we 
will show later, the initial condition will have important
influences on the approach to equilibrium and the final
entropy production.

{}For $\beta<1$, $\tau\to\infty$ corresponds to $\chi\to\infty$.
In this limit, the r.h.s. of  Eq.~(\ref{eq:solu5}) is
a finite number, $e^{\chi_0}(Y/\sinh Y-\pi/4)$. Since
the integrand on the l.h.s. of Eq.~(\ref{eq:solu5}) has
an exponential factor, a finite integral must require
\begin{equation}
  \frac{d}{d\chi'}
  [G(\chi')h((\chi'/\chi)^{\frac{1}{1-\beta}})]_{\chi'=\chi}=0\;,
  \;\;\; \chi\to\infty \; .
\end{equation}
Using $h(1)=1$ and $h'(1)=1/3$, one can find the solution to  the
above equation,
\begin{equation}
  G\propto \chi^{\frac{-1}{3(1-\beta)}} \propto (\tau_0/\tau)^{1/3}\; ,
  \;\;\; \tau/\tau_0\to\infty \; ,
\end{equation}
which corresponds to the one-dimensional hydrodynamic limit,
$\epsilon\propto (\tau_0/\tau)^{4/3}$. Therefore, thermal equilibrium
will eventually be established for $\beta<1$.

{}For $\beta>1$, $\tau\to\infty$ limit corresponds to $\chi\to - 0$.
Using $h(0)=\pi/4$, one has from Eq.~(\ref{eq:solu4})
\begin{equation}
  G(\chi)=e^{\chi_0}-\frac{\pi}{4}\int_0^{\chi_0}d\chi'G(\chi')
  ={\rm const.}\; ,\;\;\; \chi\to -0 \;.
\end{equation}
This corresponds to the free-streaming limit, $\epsilon\propto \tau_0/\tau$.
Therefore, thermal equilibrium will never be achieved if $\beta>1$.

The special case $\beta=1$ was studied by Gavin \cite{GAV91} who
found that the system will also reach an asymptotic state lying in 
between free-streaming and hydrodynamic limit depending on the value 
of the prefactor $\theta_0$ in Eq.~(\ref{eq:relax2}). In the
asymptotic state, $G(\tau/\tau_0)\propto (\tau_0/\tau)^{\gamma}$
with $0\leq\gamma<1/3$. Only for a very large $\theta_0/\tau_0 \gg 1$, 
does the system approach to the hydrodynamic limit with
$\gamma \approx 1/3-16\tau_0/45\theta_0$, which is very close
in form to the case of $\beta\la 1$ as noted by Gavin in Ref.~\cite{GAV91}.

\begin{figure}
\centerline{\psfig{figure=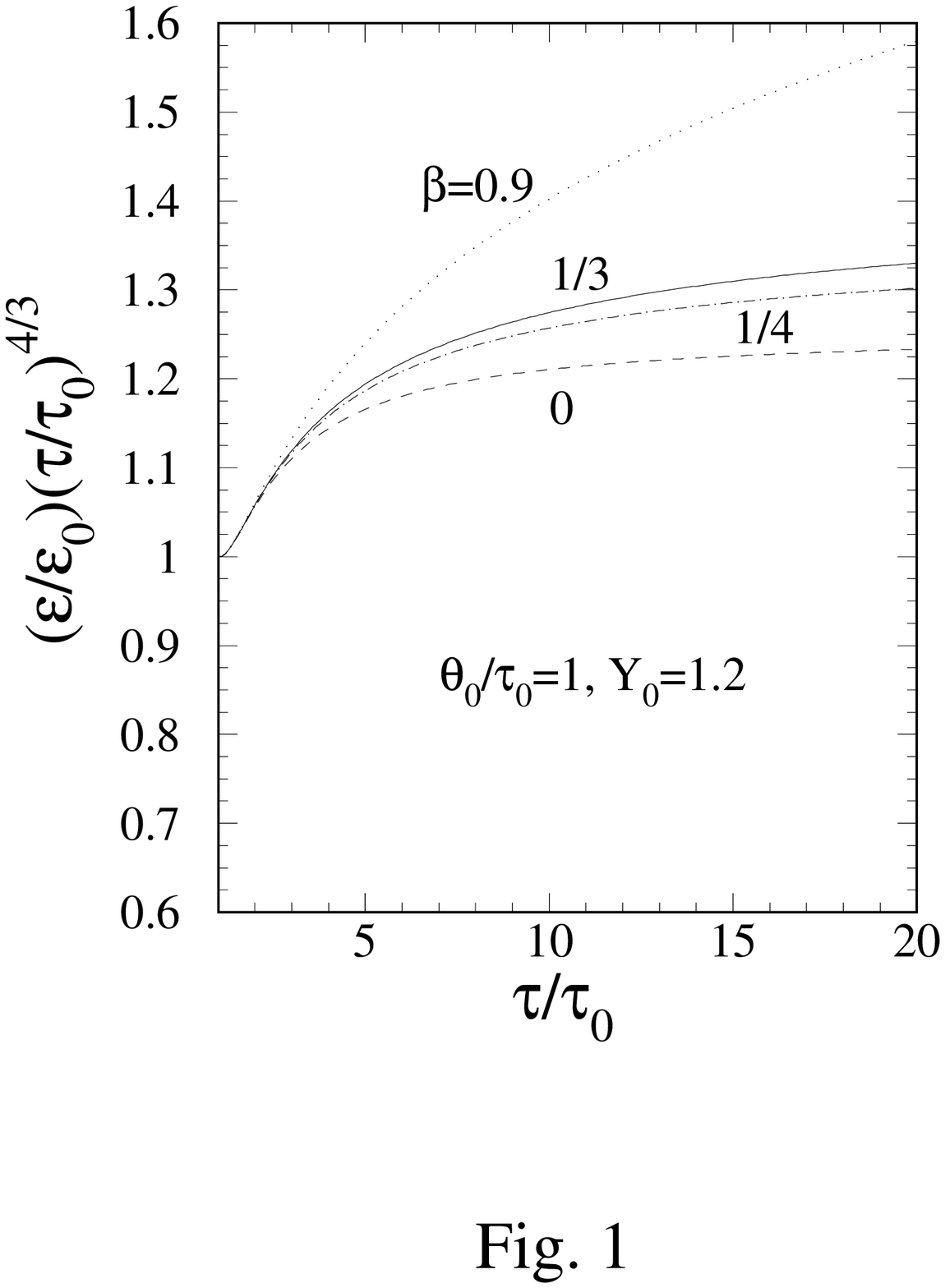,width=2.5in,height=3in}}
\caption{Time evolution of $G(\tau/\tau_0)(\tau/\tau_0)^{1/3}
    =(\epsilon/\epsilon_0)(\tau/\tau_0)^{4/3}$ according to the 
    solution to Boltzmann equation, for different time dependence
    of the relaxation time $\theta=\theta_0(\tau/\tau_0)^{\beta}$,
    with $\theta_0/\tau_0=1$, $\beta=0.9$ (dotted), 1/3 
    (solid), 1/4 (dot-dashed)and 0 (dashed line). $Y_0=1.2$ is the
    width of the initial rapidity distribution. }
\label{fig1}
\end{figure}

\begin{figure}
\centerline{\psfig{figure=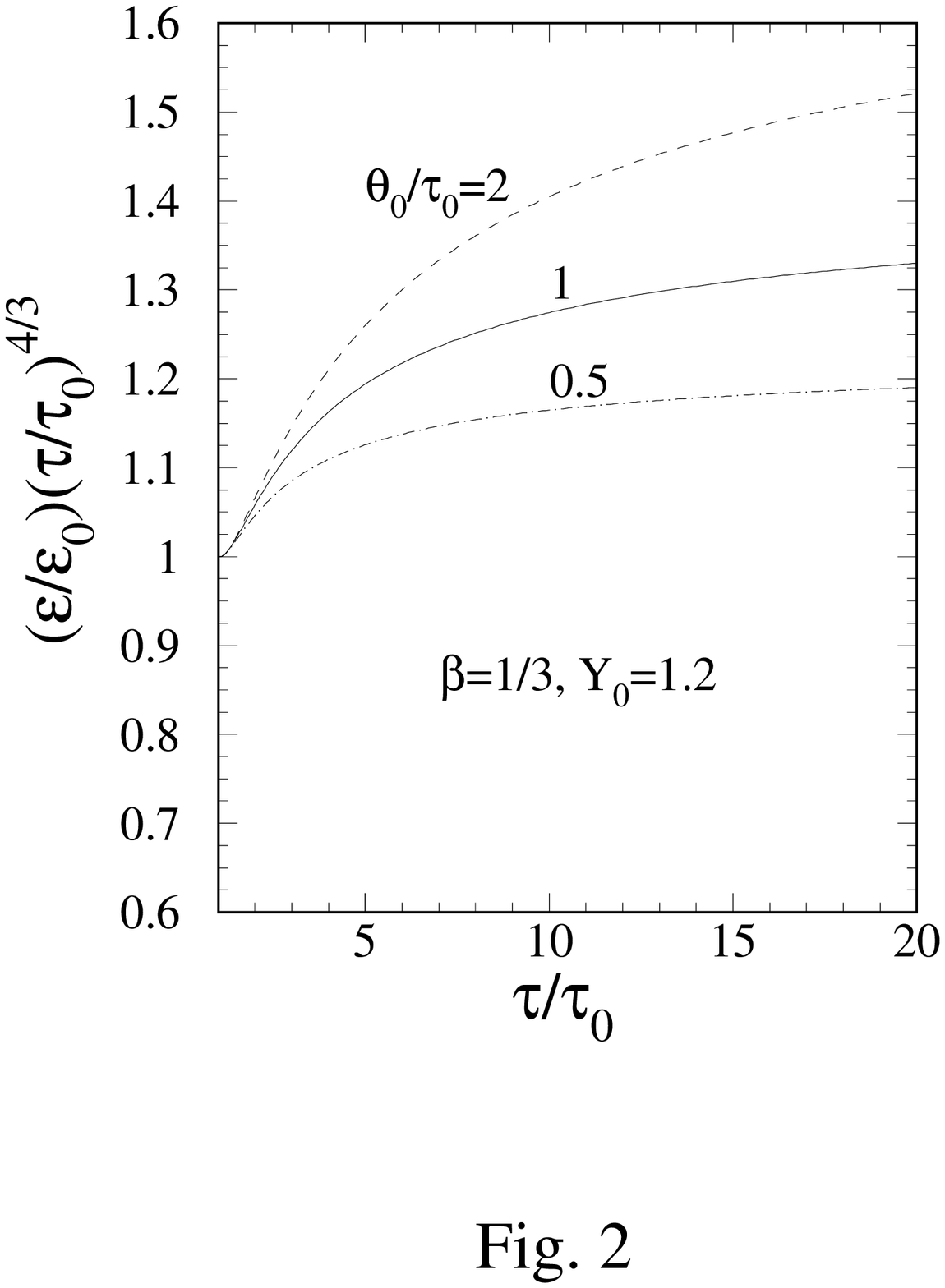,width=2.5in,height=3in}}
  \caption{Same as Fig.~(\protect{\ref{fig1}}), expect for
    $\beta=1/3$, $\theta_0/\tau_0=2$ (dashed), 1 (solid) and 0.5 
    (dot-dashed line).}
\label{fig2}
\end{figure}

Shown in Figs.~\ref{fig1} and \ref{fig2} are the numerical
solutions of Eq.~(\ref{eq:solu4}) for different relaxation times.
{}For a clear presentation, we plot 
$(\tau/\tau_0)^{1/3}G(\tau/\tau_0)=(\tau/\tau_0)^{4/3}(\epsilon/\epsilon_0)$
as a function of $\tau/\tau_0$ for different values of $\beta$ and
$\theta_0/\tau_0$. In these two plots, we have chosen $Y=1.2$ 
which corresponds to an isotropic initial distribution in momentum space.
If the system undergoes free streaming, i.e., $G(\tau/\tau_0)={\rm const.}$,
the plotted quantity should increase with $\tau/\tau_0$ with a
power of $1/3$. Thus we see from Figs.~\ref{fig1} and \ref{fig2}
that a system must undergo free streaming for a period of time
before it approaches the equilibrium limit when parton interactions
eventually balance the expansion. The duration of such
a period, which we can define as the thermalization time $\tau_{\rm th}$
is determined by the relaxation time and its time dependence.
For fixed $\beta$, the thermalization time $\tau_{\rm th}$ is
approximately proportional to $\theta_0$. From Fig.~\ref{fig2},
we can estimate that $\tau_{\rm th}\simeq 5 \theta_0$ for
$\beta=1/3$ if we consider that thermalization is reached
when $G(\tau/\tau_0)$ is about 10\% from its hydrodynamical limit. 
Assuming one scattering is sufficient to thermalize the 
system \cite{Shuryak} is therefore a serious underestimate.
When $\beta$ is close to 1, the system approaches the hydrodynamical
limit very slowly as seen in Fig.~\ref{fig1}. However, the
hydrodynamical limit is still achieved as long as $\beta<1$, unlike 
when $\beta\geq 1$. When the thermalization time is very long,
one must also consider whether 3-dimensional expansion and/or 
hadronization occur earlier.

\section{Memory effect}

\begin{figure}
\centerline{\psfig{figure=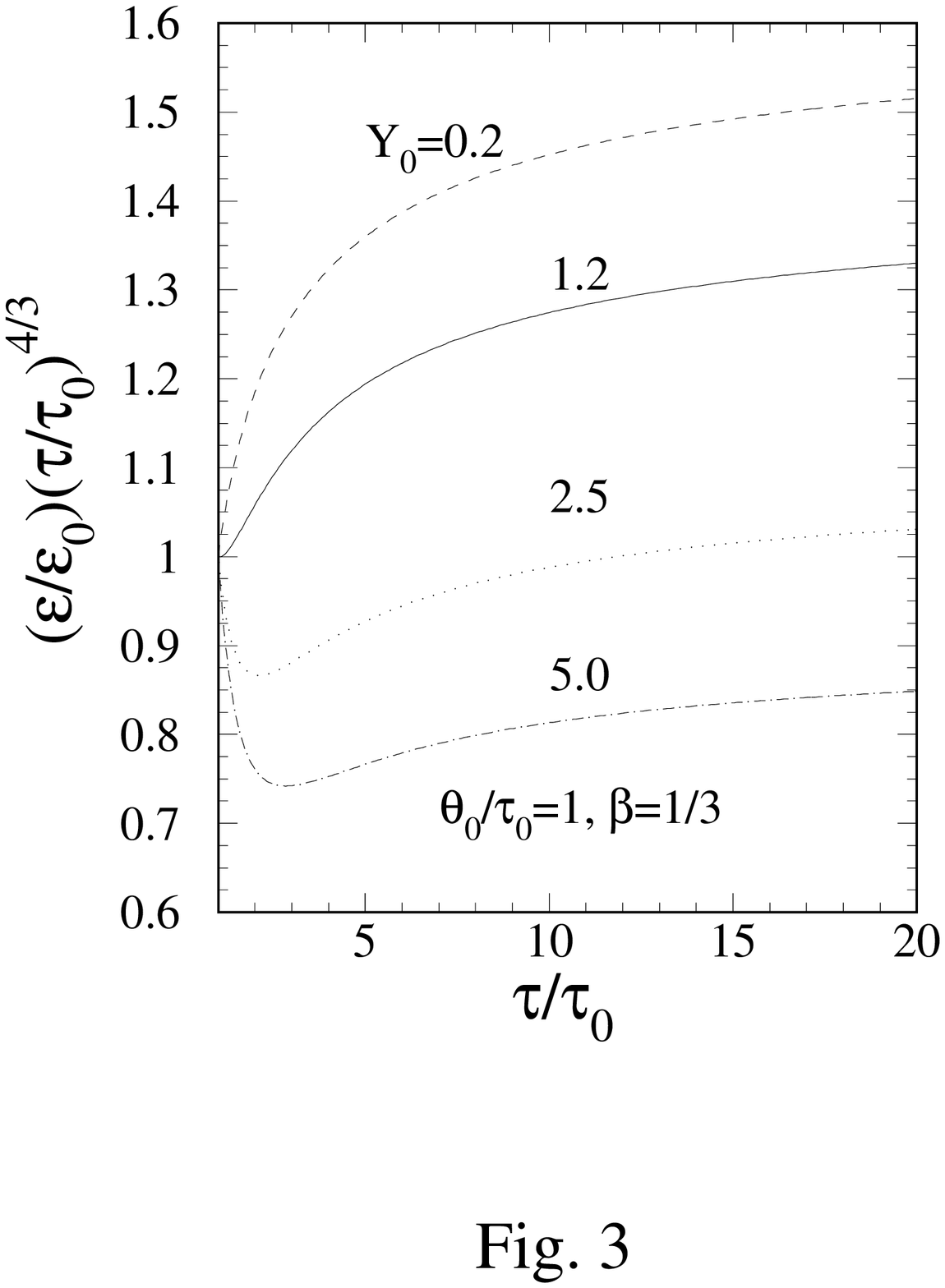,width=2.5in,height=3in}}
  \caption{Same as Fig.~(\protect{\ref{fig1}}), except for
    $\beta=1/3$, $\theta_0/\tau_0=1$, $Y_0=0.2$ (dashed),
    1.2 (solid), 2.5 (dotted) and 5.0 (dot-dashed line).}
\label{fig3}
\end{figure}

Let us now discuss how the initial condition $f_0$ influences
the thermalization process and the final entropy production.
In Fig.~\ref{fig3} the time evolution of 
$(\tau/\tau_0)^{1/3}G(\tau/\tau_0)$ is shown for different values
of $Y$ which characterizes the initial distribution of partons
in the phase space. For large values of $Y$, we notice that
the system initially expands even faster than the ideal 
hydrodynamical case and then turns over, approaching the
hydrodynamical limit. To understand this, let us take
the first moment of Eq.~(\ref{eq:boltz4}) with respect to
the single parton energy. We then have, by energy and momentum
conservation,
\begin{equation}
  \frac{d\epsilon}{d\tau}+\frac{\epsilon+P_L}{\tau}=0\;, 
  \label{eq:hydrol}
\end{equation}
where the energy density is defined by Eq.~(\ref{eq:solu3}) and
\begin{equation}
  P_L=\nu_g\int \frac{d^3p}{(2\pi)^3} \frac{p_z^2}{E} f(p)
\end{equation}
is the longitudinal pressure. The solution to the Boltzmann
equation can be parametrized as $P_L(\tau)=\gamma(\tau)\epsilon(\tau)$
with $0<\gamma<1$. We have then from Eqs.~(\ref{eq:hydrol}) and (\ref{eq:G}),
\begin{equation}
  G(\tau)=e^{-\int_{\tau_0}^{\tau}d\tau' 
    \frac{\gamma(\tau')}{\tau'}}\; .
\end{equation}
Thus, $\gamma(\tau)$ characterizes the thermalization of
the system (or equal partition in longitudinal and transverse
direction). In the hydrodynamical limit, $\gamma=1/3$, while
free-streaming corresponds to $\gamma=0$. 
One can
expect that the initial evolution of the system near
$\tau/\tau_0\sim 1$ should be determined by the value
of $\gamma_0=\gamma(\tau_0)$,
\begin{equation}
  G(\tau)\simeq (\tau_0/\tau)^{\gamma_0}\; , 
  \;\;\; \tau/\tau_0\sim 1 \; .
\end{equation}

Using the initial distribution in Eq.~(\ref{eq:init1}), we have
\begin{equation}
  \gamma_0=\frac{P_L(\tau_0)}{\epsilon_0}
  =1-\frac{\arctan(\sinh Y)}{\sinh Y} \; .
\end{equation}
An isotropic situation corresponds to $\arctan(\sinh Y_0)/\sinh Y_0=2/3$
or $Y_0=1.167$ which is very close to the value we obtained
by requiring $H_Y(0)=\pi/4$.  For $Y<Y_0$, $\gamma_0<1/3$,
the system starts its evolution more like free streaming as
we have noticed in our numerical solutions. The extreme case 
is the Bjorken scaling ansatz, $Y=0$, $\gamma_0=0$,
which corresponds exactly to free streaming. For $Y>Y_0$, 
$\gamma_0>1/3$,
the system will initially expand in the longitudinal direction
even faster than a thermal expansion, as also demonstrated
in our numerical solutions. Physically, this is caused by the
higher longitudinal pressure built up by the large amount of
partons which are distributed over a large range of rapidity
in the local frame. One can check that the expansion in this 
early stage is still dominated by free-streaming. However,
working against such high pressure costs energy thus leading to
less entropy production as we now show in the following. 

\begin{figure}
\centerline{\psfig{figure=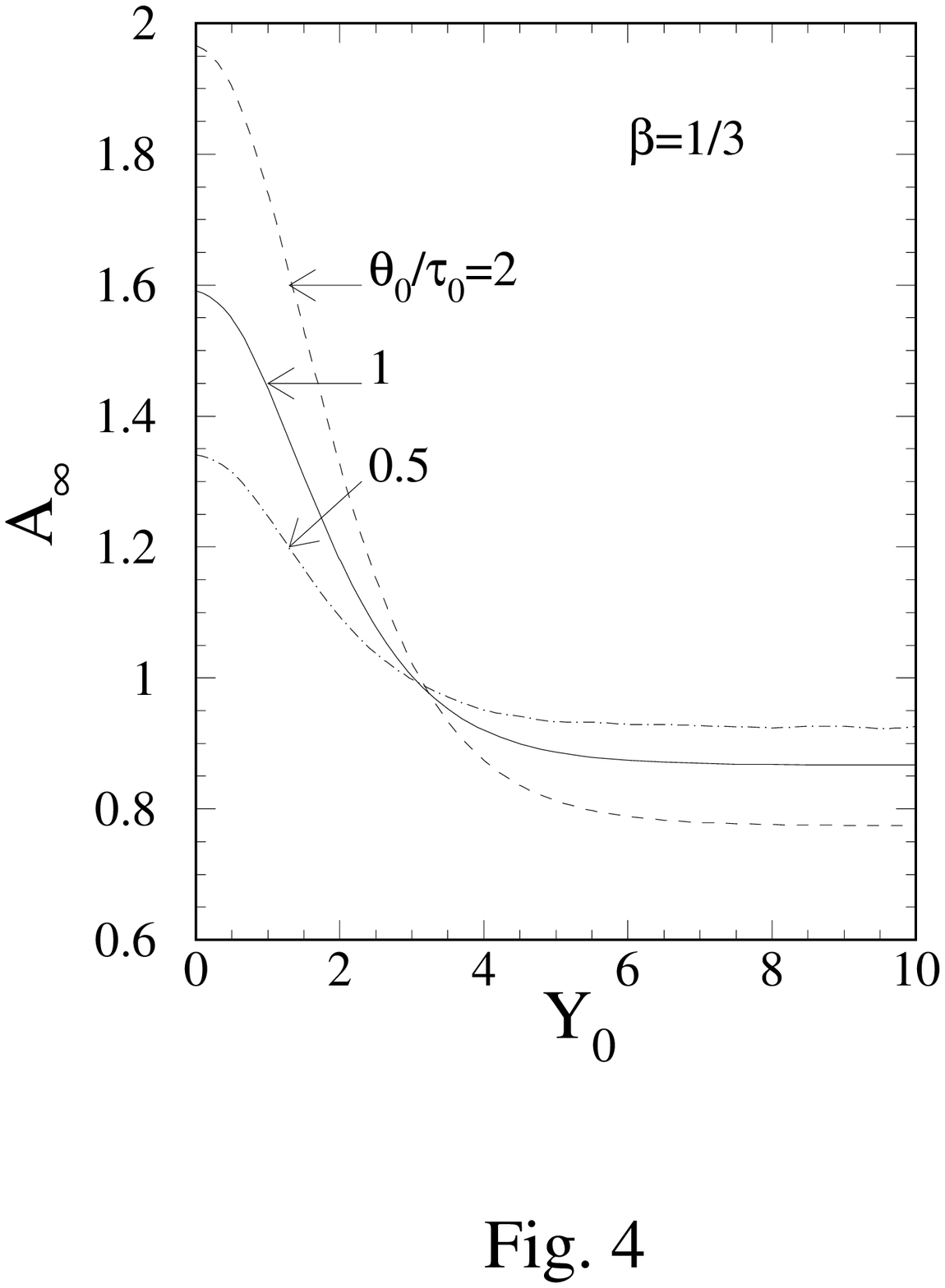,width=2.5in,height=3in}}
  \caption{The prefactor, ${\cal A}_{\infty}
    =(\epsilon/\epsilon_0)(\tau/\tau_0)^{4/3} (\tau/\tau_0\to\infty)$,
    as a function of the $Y_0$, the width of the initial rapidity
    distribution, for different relaxation times 
    $\theta=\theta_0(\tau/\tau_0)^{\beta}$ ($\beta=1/3$), 
    $\theta_0/\tau_0=2$ (dashed), 1 (solid), and 0.5 (dot-dashed line).}
\label{fig4}
\end{figure}

Since the system will eventually approach the thermal equilibrium
limit when $\beta<1$, we can define a prefactor ${\cal A}_{\infty}$ by
\begin{equation}
  G(\tau/\tau_0)={\cal A}_{\infty}(\tau_0/\tau)^{1/3}\; ,
  \;\;\; \tau/\tau_0\to\infty \;.
\end{equation}
This prefactor in general can only be calculated numerically and
will depend on the relaxation time as well as the initial 
condition as we have seen in Fig.~\ref{fig3}.
Using the above expression, one can calculate the final 
entropy of the system per comoving volume \cite{HHXW95}
\begin{equation}
  \tau s_f=\frac{4}{3}\frac{\tau_0\epsilon_0}{T_0}
  \frac{T_0}{T}G(\tau/\tau_0)=\tau_0\frac{4}{3}\frac{\epsilon_0}{T_0}
  {\cal A}_{\infty} \; . \label{eq:entropy}
\end{equation}
which is exactly proportional to ${\cal A}_{\infty}$. Here $T_0$
is only a parameter in the time dependence of the final
temperature, $T=T_0(\tau_0/\tau)^{1/3}$. Since the initial
system is not in equilibrium especially for large values of $Y$,
$(4/3)\epsilon_0/T_0$ should not be considered as the initial
entropy density. Therefore, ${\cal A}_{\infty}$ is not the 
absolute increase of the total entropy over its initial value,
except when the initial distribution is an equilibrium one. 
Nevertheless, ${\cal A}_{\infty}$ still carries a lot of information 
about the thermalization process of the system. We plot this prefactor 
in Fig.~\ref{fig4} as a function of $Y$ for $\beta=1/3$ but 
for different values of $\theta_0/\tau_0$. It is clear that 
there is less entropy production for larger values of $Y$,
since the system has to work against increasingly high
longitudinal pressure thus converting its kinetic energy to
expansion energy. 
The system has the maximum entropy production when the initial
distribution is that of the Bjorken scaling ansatz, $Y=0$.
In this case, one can calculate \cite{HHXW95} that the entropy
production increases with the relaxation time like
${\cal A}_{\infty}\propto (\theta_0/\tau_0)^{1/4(1-\beta)}$.
For very large values of $Y$, $ {\cal A}_{\infty}$ decreases
slightly with $\theta_0/\tau_0$ as indicated in Fig.~\ref{fig4},
since the system has to spend longer time work against the
extraordinaryly high longitudinal pressure. It is the competition
between long thermalization time (thus entropy production)
and long duration of work against high pressure that reverses
the $\theta_0/\tau_0$ dependence of entropy production
${\cal A}_{\infty}$.

\section{Conclusion and outlook}

In this paper, we investigated the thermalization process
in a one-dimensional expanding parton plasma within the
framework of the Boltzmann equation. In particular, we
have studied the time dependence of the relaxation time and
its influence on the thermalization. If the time dependence
is weaker than a linear form, we find that the thermal
equilibrium limit will eventually be reached. For a time dependence
stronger than the linear one, the system will never thermalize,
only leading to a free-stream limit. For an exact linear
time dependence, the system will reach an asymptotic
state between free-streaming and thermal equilibrium.
We find that the thermalization process also depends on the
initial condition of the system. The deviation of the initial
momentum distribution from an isotropic one in the longitudinal 
direction determines the initial approach to thermal equilibrium.
This initial approach will then carry its inertia throughout
the whole thermalization process. This ``memory effect'' can
be seen from the dependence of the final total entropy production
on the initial momentum distributions.

Using perturbative QCD at finite temperature in the transport
theory, we have calculated the relaxation time as a result of
parton scatterings. We have pointed out the important differences
between parton damping rates and thermalization times.
To regularize the singular behavior of the parton
scattering cross sections, we have used a full gluon propagator
which includes the resummation of an infinite number of hot thermal
loops. Because of the singular behavior of the parton scattering
matrix elements, the resultant relaxation time depends sensitively
on the Debye screening mass $q_D=gT$. In an expanding parton
gas, the temperature decreases with time and so does the Debye
screening mass, thus leading to an increasing transport cross
section. This then compensates the decrease of the parton density
and gives us a relaxation time with a weak time dependence,
$\theta\propto (\tau/\tau_0)^{1/3}$.
However, if one introduces a fixed momentum cut-off to parton
scatterings as in most of numerical simulations, one will
introduce an extra logarithmic time dependence which will
slow down the thermalization process.

Although we have demonstrated the necessity of the
inclusion of Debye screening and Landau damping in the study
of the parton thermalization in an expanding system, it is
not clear to us how to incorporate them into numerical simulations
such as parton cascade models. One can include the Debye screening
semiclassically \cite{BMW} by brutal force. However, one immediate 
problem we have to solve is how to avoid double counting.

Since we used the full propagator which includes many thermal 
loops, we have effectively included
multiple particle scatterings. One can easily see this by
expanding the full propagator in terms of the bare propagators
at zero temperature and the self-energy from thermal loops.
The contribution from the real part of the self-energy (mainly
Debye screening) corresponds to multiple particle scatterings
in the thermal bath. One can in principle include particle 
radiation and absorption by considering the thermal loop
corrections to the full vertices. In doing so, one can
automatically avoid both the infrared and collinear
divergencies \cite{WELD94} which one normally regularizes
by resorting to two additional cut-offs \cite{LXES}.

As we have demonstrated in Appendix B, the imaginary part
of the self-energy also contributes to the effective
parton scattering in a parton gas. In fact, the imaginary
part, which is responsible for Landau damping, is necessary
to regularize the transverse interaction since there is no
magnetic screening in QCD. The contribution from this
imaginary part to an effective two-parton scattering 
corresponds to independent scatterings of the two partons
off particles in the thermal bath. This can easily cause
double counting in a parton cascade model because independent
parton scatterings are also simulated  over the
volume of the system. One might be able to avoid this problem
by introducing a length scale in the order of $1/gT$, within
which  multiple particle scatterings are not allowed
and only {\it one} effective two-parton scattering (with Debye 
screening and Landau damping) is permitted.

\section*{Acknowledgement}
This work was supported by the Director, Office of Energy Research,
Office of High Energy and Nuclear Physics, Division of Nuclear Physics,
of the U.S. Department of Energy under Contract DE-AC03-76SF00098, 
and the Danish Natural Science Research Council. Discussions with
V.~Koch are gratefully acknowledged.

\section*{Appendices}

\appendix

\section{Effective matrix elements of forward parton scatterings}

In this appendix we separate the QCD interactions
into longitudinal and transverse parts with inclusion of screening 
in these two parts and show the simplifications that appear in 
the limit of forward scatterings or small momentum transfer interactions.
We adopt the same convention as in Ref.~\cite{WELD82} and
split a vector $Q_{\mu}$ into its components  parallel and 
orthogonal to the flow velocity $u_{\mu}$, ($u^\mu u_\mu=1$), such that
\begin{equation}
  \omega=Q\!\cdot\!u , \;\;
  \tilde{Q}_{\mu}=Q_{\mu}-u_{\mu}(Q\!\cdot\!u)\; .
\end{equation}
We can then denote a vector by $Q=[\omega, {\bf q}]$, with
$Q^2=\omega^2-q^2$ and
$\tilde{Q}^2=-q^2$. In the local frame, where the flow velocity 
is $u=(1,{\bf 0})$, $\omega$ and ${\bf q}$ become the time and 
spatial components of the vector. Similarly, one also defines
a tensor orthogonal to $u_{\mu}$,
\begin{equation}
  \tilde{g}_{\mu\nu}=g_{\mu\nu}-u_{\mu}u_{\nu}\; .
\end{equation}
The full gluon propagator with momentum $Q$
is obtained from the vacuum polarization by using Dyson's equation,
\begin{equation}
\Delta^{\mu \nu}=\frac{{\cal P}_T^{\mu\nu}}{-Q^2+\Pi_T}
+\frac{{\cal P}_L^{\mu\nu}}{-Q^2+\Pi_L}+(\alpha-1)\frac{Q^{\mu}Q^{\nu}}{Q^4},
\end{equation}
where $\alpha$ is a gauge fixing parameter,  the longitudinal
${\cal P}_L^{\mu\nu}$, and transverse tensor ${\cal P}_T^{\mu\nu}$ 
are defined as
\begin{eqnarray}
  {\cal P}_L^{\mu\nu}&=&\frac{-1}{Q^2q^2}(\omega Q^{\mu}-Q^2u^{\mu})
  (\omega Q^{\nu}-Q^2u^{\nu})\, , \\
  {\cal P}_T^{\mu\nu}&=&\tilde{g}^{\mu\nu}
  +\frac{\tilde{Q}^{\mu}\tilde{Q}^{\nu}}{q^2}\, ,
\end{eqnarray}
which are orthogonal to $Q^{\mu}$ and also to each other, i.e.,
\begin{equation}
Q_{\mu}{\cal P}_L^{\mu\nu}=Q_{\mu}{\cal P}_L^{\mu\nu}
={\cal P}^{\mu}_{L\,\nu}{\cal P}_T^{\nu\rho}=0 .
\end{equation}
In additon, one also has 
${\cal P}^{\mu\rho}{\cal P}_{\rho\nu}={\cal P}^\mu_\nu$.
The free gluon propagator at zero temperature is in this case
\begin{equation}
D^{\mu\nu}=\left(-g^{\mu\nu}+\alpha\frac{Q_{\mu}Q_{\nu}}{Q^2}\right)
\frac{1}{Q^2}\, .
\end{equation}
If we choose Feynman gauge ($\alpha=1$), the full propagator also
satisfies $Q_{\mu}\Delta^{\mu\nu}=0$. The transverse and longitudinal
self-energies are \cite{WELD82}
\begin{equation}
\Pi_L(Q)=(1-x^2)\pi_L(x)\,\,\, , \Pi_T(Q)=\pi_T(x) \label{eq:self3}
\end{equation}
with the scaled self-energies, $\pi_L(x)$ and $\pi_T(x)$, 
given in Eqs.~(\ref{eq:self1}) and (\ref{eq:self2}), where $x=\omega/q$. 

Using the full gluon propagator, $\Delta^{\mu\nu}$, we can obtain
the effective matrix element of quark scatterings
$q_i(P_1)+q_j(P_2)\to q_i(P_3)+q_j(P_4)$ ($i\neq j$), 
\begin{eqnarray}
|{\cal M}_{qq}|^2&=&C_{qq}g^4 4\{4|P_1\!\cdot\!\Delta\!\cdot\! P_2|^2
-2(P_1\!\cdot\! P_3)(P_2\!\cdot\!\Delta\!\cdot\!\Delta^*\!\cdot\! P_4) 
\nonumber \\
 &\,& -2(P_2\!\cdot\! P_4)(P_1\!\cdot\!\Delta\!\cdot\!\Delta^*\!\cdot\! P_3)
+(P_1\!\cdot\! P_3)(P_2\!\cdot\! P_4)|\Delta|^2 \}\; , \label{eq:matrix2}
\end{eqnarray}
where $C_{qq}=(N_c^2-1)/4N_c^2=2/9$ is the color factor and
\begin{eqnarray}
  \Delta^{\mu\rho}\Delta_{\rho}^{*\,\nu}&=&
\frac{{\cal P}_T^{\mu\nu}}{|Q^2-\Pi_T|^2}
+\frac{{\cal P}_L^{\mu\nu}}{|Q^2-\Pi_L|^2}\; ,\\
|\Delta|^2=\Delta^{\mu\nu}\Delta^*_{\nu\mu}&=&
\frac{2}{|Q^2-\Pi_T|^2}
+\frac{1}{|Q^2-\Pi_L|^2}\; .
\end{eqnarray}
For small angle scatterings, 
$-Q^2/2=P_1\!\cdot\! P_3=P_2\!\cdot\! P_4\ll P_1\!\cdot\! P_2
=P_3\!\cdot\! P_4$,
only the first term, corresponding to t-channel scattering, 
in Eq.~(\ref{eq:matrix2}) is dominant.
Furthermore, $\omega=xq\approx{\bf v_1}\!\cdot\!{\bf q}
\approx {\bf v_2}\!\cdot\!{\bf q}$
from the energy and momentum conservation. In this approximation,
one can verify that
\begin{eqnarray}
P_1\!\cdot\! {\cal P}_L\!\cdot\! P_2&=&
-E_1E_2({\bf v}_1\!\cdot\!{\bf v}_2-x^2)\;\; ,\\
P_1\!\cdot\! {\cal P}_T\!\cdot\! P_2&=&E_1E_2(1-x^2) \;\; .
\end{eqnarray}

Define $\cos\phi=
({\bf v}_1\times\hat{\bf q})\!\cdot\!({\bf v}_1\times\hat{\bf q})$,
we can express ${\bf v}_1\!\cdot\!{\bf v}_2$ as
\begin{eqnarray}
  {\bf v}_1\!\cdot\!{\bf v}_2&=&
 ({\bf v}_1\!\cdot\!\hat{\bf q})({\bf v}_2\!\cdot\!\hat{\bf q}) 
 +\cos\phi\sqrt{1-({\bf v}_1\!\cdot\!\hat{\bf q})^2}
 \sqrt{1-({\bf v}_2\!\cdot\!\hat{\bf q})^2} \nonumber \\
 &=&x^2+(1-x^2)\cos\phi \; .
\end{eqnarray}
We have then
\begin{equation}
|{\cal M}_{qq}|^2\approx C_{qq} g^4 16 (E_1E_2)^2
\left|\frac{1-x^2}{\omega^2-q^2-\Pi_L}
-\frac{(1-x^2)\cos\phi}{\omega^2-q^2-\Pi_T}\right|^2 \; .
\end{equation}
The matrix elements for gluon-quark and gluon-gluon scatterings
are similar in the small angle approximation, except the color
factors, $C_{gq}=1/2$ and $C_{gg}=N_c^2/(N_c^2-1)=9/8$.
For scatterings of identical particles, one should also
multiply a factor of 2 to take into account of the equal
contributions of $t$ and $u$-channel scatterings.
Using $E_3\approx E_1$, $E_4\approx E_2$, and Eq.~(\ref{eq:self3})
one arrives at Eq.~(\ref{eq:matrix1}).

\section{Calculation of the relaxation time}
For a system near thermal equilibrium, the energy-momentum
tensor is given to the zeroth order of the deviation $\delta f$
by 
\begin{equation}
T^{\mu\nu}_{(0)}=(\epsilon+P)u^{\mu}u^{\nu}-g^{\mu\nu}P\; ,
\end{equation}
where $\epsilon$ is the energy density and $P$ the pressure.
One can also split the derivative $\partial_{\mu}$ into
components parallel and orthogonal to the flow velocity $u^{\mu}$, 
\begin{equation}
\partial_{\mu}=u_{\mu} D+\tilde{\partial}_{\mu},
\end{equation}
where $D=u\!\cdot\!\partial$ and $\tilde{\partial}_{\mu}
=\partial_{\mu}-u_{\mu}D$.
The energy-momentum conservation $\partial_{\nu}T^{\nu\mu}_{(0)}=0$
can be rewritten as
\begin{eqnarray}
  D\epsilon+(\epsilon+P)\tilde{\partial}\!\cdot\!u&=&0\; , \nonumber \\
  (\epsilon+P)Du^{\mu}-\tilde{\partial}^{\mu}P&=&0\; . \label{eq:hydro1}
\end{eqnarray}
Conservation of entropy 
requires that the entropy density $s=(\epsilon+P)/T$ fulfills
$\partial_{\mu}(su_{\mu})=Ds+s\tilde{\partial}\!\cdot\!u=0$. Thus we
obtain 
\begin{equation}
  \frac{DT}{T}=-\frac{\partial P}{\partial \epsilon} 
  \tilde{\partial}\!\cdot\!u \; , \label{eq:hydro2}
\end{equation}
where $\partial P/\partial \epsilon=1/3$ for an ideal gluon gas.

We assume the space-time variation of the distribution $f$ is
small. To the leading order in a gradient expansion in the relaxation
time approximation, we have
\begin{equation}
  \delta f=-\frac{\theta(p)}{p\!\cdot\!u} p\!\cdot\!\partial f^{\rm eq}
  =\frac{\theta(p)}{p\!\cdot\!u} f^{\rm eq}(1\pm f^{\rm eq})
  p\!\cdot\!\partial \left(\frac{p\!\cdot\!u}{T}\right) \; .
\end{equation}
Using the hydrodynamical relations, Eqs.~(\ref{eq:hydro1}) and
(\ref{eq:hydro2}), one can rewrite the above as
\begin{equation}
  \frac{\delta f}{f^{\rm eq}(1\pm f^{\rm eq})}
  = \frac{\theta(p)}{T} p^{\mu}X_{\mu}
  +\frac{\theta(p)}{T}\Sigma^{\mu\nu} U_{\mu\nu}\; ,
\end{equation}
where $U_{\mu\nu}=(1/2)(\tilde{\partial}_{\mu}u_{\nu}
+\tilde{\partial}_{\nu}u_{\mu})$, 
$X_{\mu}=\tilde{\partial}_{\mu}P/(\epsilon+P)-\tilde{\partial}_{\mu}T/T$
which vanishes for an ideal quark and gluon gas and $\Sigma_{\mu\nu}$ is
another tensor orthogonal to the flow velocity,
\begin{equation}
\Sigma^{\mu\nu}(p)=(p\!\cdot\!u)\left[
  \frac{\tilde{p}^{\mu}\tilde{p}^{\nu}}{(p\!\cdot\!u)^2}
  +\frac{1}{3}\tilde{g}^{\mu\nu}\right] \; .
\end{equation} 
To leading order in the gradient expansion, the Boltzmann
equation becomes
\begin{eqnarray}
  f^{\rm eq}_1(1\pm f^{\rm eq}_1)\Sigma^{\mu\nu}(p_1)&=&
  \nu_g\int dp_2dp_3 dp_4 \left[ \theta(p_1)\Sigma_1^{\mu\nu}
  +\theta(p_2)\Sigma_2^{\mu\nu}-\theta(p_3)\Sigma_3^{\mu\nu}
  -\theta(p_4)\Sigma_4^{\mu\nu}\right] \\ \nonumber
  &\ &f^{\rm eq}_1f^{\rm eq}_2
  (1\pm f^{\rm eq}_3)(1\pm f^{\rm eq}_4)
  \frac{1}{2}|M_{12\to 34}|^2 (2\pi)^4\delta^4(P_1+P_2-P_3-P_4)\; .
  \label{eq:boltz5}
\end{eqnarray}
In general, the relaxation time is momentum dependent and
it, or the deviation from local equilibrium, must be determined by solving
the Boltzman equation. As a first approximation, we assume
$\theta$ is independent of the momentum $p$ (one can also
assume $\theta\propto p$ which is found to be a 
good approximation in calculations of viscosities \cite{HH94}).
Multiplying both sides of the above equation by $\Sigma^{\mu\nu}(p_1)$
and integrating over $dp_1$, we have then
\begin{equation}
  \theta=\frac{\int dp 
    f^{\rm eq}(1\pm f^{\rm eq})\Sigma^2(p)}
  {{\cal L}[\Sigma]}\; ,\\
\end{equation}
where
\begin{eqnarray}
  {\cal L}[\Sigma]&=&\frac{\nu_g}{(2\pi)^8}
  \int d^3p_1 f^{\rm eq}_1(1\pm f^{\rm eq}_3)
  \int d^3p_2 f^{\rm eq}_2(1\pm f^{\rm eq}_4)
  \int d^3q\int d\omega 
  \frac{1}{4}(\Sigma_1+\Sigma_2 -\Sigma_3-\Sigma_4)^2 
  \nonumber \\
  &\ & C_{gg} g^4 \left|\frac{1}{q^2+\pi_L(x)}
  -\frac{(1-x^2)\cos\phi}
  {q^2(1-x^2)+\pi_T(x)}\right|^2
  \delta(\omega-{\bf v_1}\!\cdot\!{\bf q})
  \delta(\omega-{\bf v_2}\!\cdot\!{\bf q})\; ,
\end{eqnarray}
where again we have made the approximation of dominance of small 
angle scatterings in the local frame, in which $u=(1,{\bf 0})$
and $\Sigma_{\mu\nu}$ only has nonvanishing spatial components,
\begin{equation}
  \Sigma_{ij}=\frac{1}{E}\left(p_ip_j-\frac{1}{3}E^2\delta_{ij}\right) \;,
  \;\; {\rm and}\;\; \Sigma^2=\frac{2}{3}E^2 \; .
\end{equation}
Using Eq.~(\ref{eq:consv}) we have
\begin{equation}
(\Sigma_1+\Sigma_2 -\Sigma_3-\Sigma_4)_{ij}\approx 
(v_2-v_1)^iq^j+(v_2-v_1)^jq^i-\omega (v_2^iv_2^j-v_1^iv_1^j)\; ,
\end{equation}
for small $q$ and $\omega$. In terms of $\cos\phi$ and $x=\omega/q$,
\begin{eqnarray}
  (\Sigma_1+\Sigma_2 -\Sigma_3-\Sigma_4)^2
  &=&2(1-{\bf v}_1\!\cdot\!{\bf v}_2)[2q^2-2\omega {\bf q}\!\cdot\!
  ({\bf v}_1+{\bf v}_2)+\omega^2(1+{\bf v}_1\!\cdot\!{\bf v}_2)] \nonumber \\
  &=&2q^2(1-\cos\phi)(1-x^2)^2(2-x^2+x^2\cos\phi) \; .
\end{eqnarray}
One can easily complete the angular integrations in ${\cal L}[\Sigma]$
similarly as in the calculation of the gluon damping rate, and has
\begin{eqnarray}
  {\cal L}[\Sigma]&=& \frac{\nu_gC_{gg}g^4}{(2\pi)^5}
  \int dp_1 p_1^2f^{\rm eq}_1(1\pm f^{\rm eq}_1)
  \int dp_2 p_2^2f^{\rm eq}_2(1\pm f^{\rm eq}_2)
  {\cal I}(q_{\rm max},q_D)\; , \\
  {\cal I}(q_{\rm max},q_D)&=&\int_0^{q^2_{\rm max}} dq^2\int_{-1}^1dx
  q^2(1-x^2)^2\left\{\frac{1-\frac{3}{4}x^2}{|q^2+\pi_L(x)|^2} 
  \right. \nonumber \\ &+&\left. 
  \frac{1}{2}(1-\frac{7}{8}x^2)\frac{(1-x^2)^2}{|q^2(1-x^2)+\pi_T(x)|^2} 
 + {\rm Re}
\left( \frac{(1-x^2)^2}{[q^2+\pi^*_L(x)][q^2(1-x^2)+\pi_T(x)]}\right) \right\}
\end{eqnarray}
After performing integrations like
\begin{equation}
  \int_0^{\infty}dp f^{\rm eq}(1+f^{\rm eq})p^n
  =n!\,\zeta(n)T^{n+1} \, ,
\end{equation}
for a Bose-Einstein distribution $f^{\rm eq}$,
where $\zeta(n)$ is the Riemann's $\zeta$-function, we obtain for the
relaxation time,
\begin{equation}
  \frac{1}{\theta}=\frac{5N_c^2}{4\pi}\frac{g^4}{(4\pi)^2} T
  {\cal I}(q_{\rm max},q_D) \; .
\end{equation}
The integral ${\cal I}(q_{\rm max},q_D)$ has no dimension and
therefore should be only a function of $q_{\rm max}/q_D$.
Numerical evaluation of the integral gives
\begin{equation}
  {\cal I}(q_{\rm max}/q_D)= 
  2.3 \ln(q_{\rm max}^2/q_D^2)-0.62+2.7\frac{q_D^2}{q_{\rm max}^2}
  + {\cal O}(q_D^4/q_{\rm max}^4)
\end{equation}
for large values of $q_{\rm max}/q_D$. Considering  Debye
screening from only gluon interactions, we have
\begin{equation}
  \frac{1}{\theta}\approx 0.92 N_c^2\alpha_s^2 T
  \left[\ln\left(\frac{1.6}{N_c\alpha_s}\right)+\cal{O}(\alpha_s)\right]\; .
\end{equation}
Similarly as we argued  after Eq.~(\ref{eq:20}), contributions
to the order $\alpha_s^3$ must be neglected since they
are not complete in our calculation.

\end{document}